%% file: main.tex
\documentclass[acmlarge,authorversion,nonacm]{acmart}

\input{macros}

\AtBeginDocument{%
  \providecommand\BibTeX{{%
    \normalfont B\kern-0.5em{\scshape i\kern-0.25em b}\kern-0.8em\TeX}}}

\begin{document}

\title{Reranking Social Media Feeds: A Practical Guide for Field Experiments}

\author{Tiziano Piccardi}
\authornote{These authors contributed equally to this work.}
\affiliation{%
  \institution{Johns Hopkins University}
  \country{USA}}
\email{piccardi@jhu.edu}

\author{Martin Saveski}
\authornotemark[1]
\affiliation{%
  \institution{University of Washington}
  \country{USA}}
\email{msaveski@uw.edu}

\author{Chenyan Jia}
\authornotemark[1]
\affiliation{%
  \institution{Northeastern University}
  \country{USA}}
\email{c.jia@northeastern.edu}

\author{Jeff Hancock}
\affiliation{%
  \institution{Stanford University}
  \country{USA}}
\email{hancockj@stanford.edu}

\author{Jeanne L. Tsai}
\affiliation{%
  \institution{Stanford University}
  \country{USA}}
\email{jltsai@stanford.edu}

\author{Michael S. Bernstein}
\affiliation{%
  \institution{Stanford University}
  \country{USA}}
\email{mbernst@stanford.edu}

\renewcommand{\shortauthors}{Piccardi et al.}

\begin{abstract}
Social media plays a central role in shaping public opinion and behavior, yet performing experiments on these platforms and, in particular, on feed algorithms is becoming increasingly challenging. This guide offers practical recommendations for researchers developing and deploying field experiments focused on real-time reranking of social media feeds. The article is organized around two contributions. First, we provide an overview of an experimental method using web browser extensions that intercepts and reranks content in real time, enabling naturalistic reranking field experiments. We then describe feed interventions and measurements that this paradigm enables on participants' actual feeds, without requiring the involvement of social media platforms. Second, we offer concrete technical recommendations for intercepting and reranking social media feeds with minimal user-facing delay, and provide an open-source implementation. This document aims to summarize lessons learned in running field experiments on social media, provide concrete implementation details, and foster the ecosystem of independent social media research. Finally, we release the source code that serves as a blueprint for implementing future feed-ranking experiments: \url{https://github.com/StanfordHCI/FeedMonitor}.

\end{abstract}

\maketitle
\section{Introduction}

\label{sec:teaser}
\begin{figure*}[t]
    \includegraphics[width=0.98\linewidth]{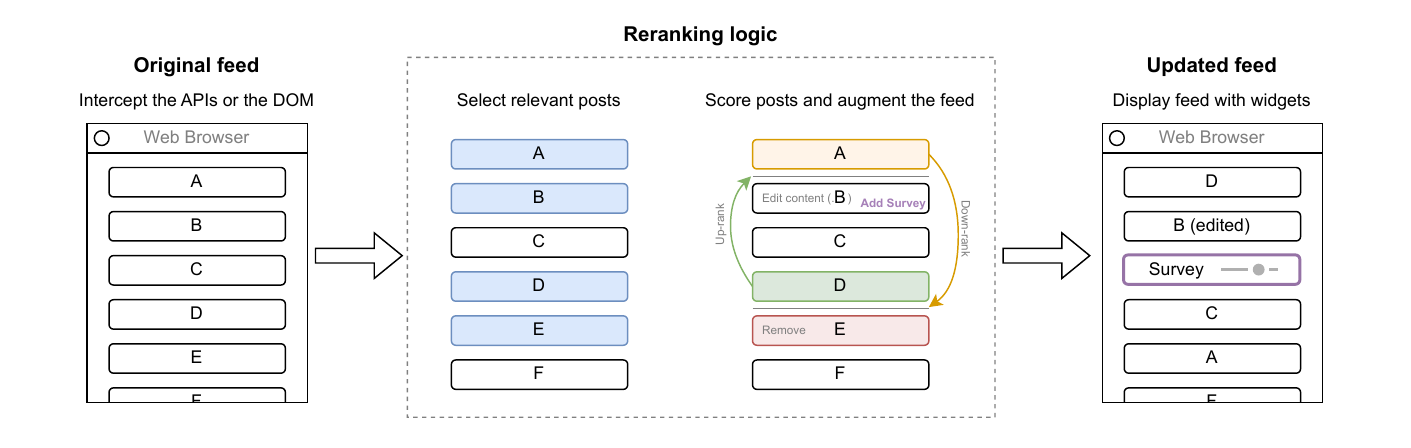}
    \caption{Diagram of the typical feed reranking steps using a browser extension. First, the extension identifies the posts that need to be rescored (optional if all posts are included). The ranking model assigns a new score to the posts, and the updated data is finally displayed in the browser. The modified feed can be enriched with custom widgets, such as surveys, to assess the intervention's impact.}
    \label{fig:extension_flow}
\end{figure*}

Social media plays a critical role in forming public opinion and influencing both individual behaviors and societal outcomes~\cite{weeks2017online}. 
Many rely on social media to follow the news and stay informed~\cite{social_media_news_use}.
Central to this media consumption behavior is the \textit{feed}, which selects and ranks the social media posts users see. It is the primary interface through which platforms direct users’ attention and users engage with content. Most major platforms, including Facebook, X (Twitter), and Reddit, rely on their feeds to provide users with a centralized view of relevant content. However, in doing so, these platforms must make algorithmic curation choices~\cite{backstrom2016serving}. In response, researchers have called for special attention to the impact of these feed algorithms, for example, the kinds of content and behaviors that they up-rank or down-rank~\cite{allen2025platform}. These design choices are of substantial interest because they may profoundly shape our experiences and beliefs. 
Of particular concern are feed algorithms that aim to maximize user engagement and inadvertently amplify undesirable content, which may exacerbate polarization~\cite{rathje2021out,kubin2021role}, misinformation~\cite{valenzuela2019paradox}, and antisocial behaviors \cite{saurwein2021automated}. %

Investigating the causal impact of the feed ranking in an ecologically valid way is challenging. Despite efforts to open-source some of the key components of the platforms' ranking algorithms~\cite{twitter_alg}, their ``black box'' nature, reinforced by the lack of data \cite{tromble2021have}, complicates attempts to evaluate their causal effects on individual users and society as a whole. While the ``logic'' behind the algorithm may be open source, the algorithms' outputs inherently depend on closed machine learning models that score aspects of each post and determine what is visible in the feed and in what order. Another challenge is the increasingly limited access to the users' post \textit{inventory}, i.e., the posts that would be considered to be shown in the users' feeds. Incomplete access to the key components of the feed—algorithms, models, and inventory—inherently limits the validity of any investigation.

This article aims to bridge this gap by offering a practical guide for conducting independent field experiments to investigate the impact of feed ranking in social media, building on our own experience in designing and deploying such interventions~\cite{piccardi2025reranking}. The guidelines focus on facilitating the design of experiments that do not require direct cooperation with the social media platform, enabling audits of feed algorithms and the development of interventions to improve the design of online social spaces. We describe how to implement experiments on the web interfaces of social media platforms using browser extensions, a feature available in all major browsers, by intercepting and customizing the network requests that populate the feed. We focus on specific interventions that this method supports, including feed reranking and content editing, and describe the relative advantages and disadvantages of each. We do not aim to exhaustively list all possible ways to implement a field experiment, but we do provide concrete building blocks for designing custom experiments.

Figure~\ref{fig:extension_flow} summarizes the main steps of a typical field experiment designed to customize the ranking of a social media feed with a browser extension. This guide focuses on designing interventions that do not affect the users' typical experience on the platform, aiming to limit latency and any interface behavior that can disrupt the user experience. To achieve this goal, the browser extension (1) captures the feed, (2) applies any experimental transformations, and then (3) inserts any custom user interface elements. First, it reads the feed to identify posts that need editing based on the experiment's objective. For example, the experiment may focus on all the content in the feed, or it may target specific posts, such as politics or misinformation. Then, the extension can apply the desired intervention to the selected posts (or newly inserted posts), such as changing their ranking, rewriting their text, or editing their social metrics. Finally, the extension can change the webpage's design or add widgets that enhance the interface and enable new user interactions. Such modifications enable researchers to design custom user interactions or assess the effect of a specific intervention by adding targeted surveys.

This article is organized as follows. In Section \ref{sec:background}, we summarize the state-of-the-art and highlight studies closely related to our work. In Section \ref{sec:interventions}, we present opportunities and challenges in changing the feed ranking with a browser extension, followed by an overview of possible approaches to measuring the outcomes of the field experiment. Finally, Section \ref{sec:implementation} provides key implementation details to develop the feed experiment, ranging from ensuring low-effort user registration to intercepting and editing the feed. We conclude with a discussion of implications. With this article, we release an open-source implementation of the proposed framework~\cite{stanfordhci_feedmonitor_2026}.

\section{Background}
\label{sec:background}
In this section, we summarize previous efforts to study social media through experiments and position this article in the current literature.
On-platform feed experiments are the gold standard for inferring the causal effects of interventions on social media, as they test interventions in the same setting where users would typically encounter them, maximizing ecological validity. An alternative to running experiments on the platform is to create experimental clones of the platform and simulate a social media environment~\cite{voggenreiter2023role,epstein2022yourfeed,jia2024embedding}. Developing a custom social media feed simulator is a meaningful approach, as it allows greater control over the experimental setup, including the posts the user sees and the features offered by the websites. This full control over the feed has significant advantages, as participants can be exposed to the same content, potentially leading to a more controlled experiment that can help measure small effects. However, the downside is that participants' online experiences are generally hard to reproduce, and the generalizability of the findings may be limited.  Participants may be exposed to a different interface and a different type of posts they typically consume, miss realistic social signals and dynamics, or be required to engage in activities not aligned with their daily habits. The requirements to engage with a custom platform may also exacerbate phenomena such as the Hawthorne effect, potentially altering participants' behavior in ways that do not fully generalize to a real-world scenario.

Previous work shows the effectiveness of field experiments on social media and that they can bring valuable findings. Some examples include investigating the effect of changing Facebook's feed ranking from engagement-based to chronological during a US election~\cite{guess2023social}, the impact and propagation of emotions~\cite{kramer2014experimental}, the amplification of political discourse~\cite{huszar2022algorithmic}, the effect of perspective-taking on political polarization~\cite{saveski2022perspective}, or using ads to reduce misinformation~\cite{lin2024reducing} and support charitable donations~\cite{Jilke2019Using}.

However, designing a field experiment with real-time intervention on social media feeds poses many technical challenges, especially when conducted without the platforms' involvement~\cite {allen2025platform,mosleh2022field,guess2021experiments}. The present work aims to support researchers with practical recommendations for designing field experiments focused on feed ranking.

Previous contributions similar in spirit to this work include practical guides on designing studies on fake news and misinformation~\cite{pennycook2021practical}, analyzing browsing data~\cite{clemm2023analysis}, designing surveys to measure political behavior online~\cite{guess2023digital}, sampling survey participants~\cite{rosenzweig2020survey}, running split tests on Facebook~\cite{orazi2020running}, and software and techniques to collect and process social media data~\cite{batrinca2015social}.

While many field experiments of interest only require changing participants' feeds, some experiments remain out of scope for this approach. For example, this approach cannot be used to perform network-based randomized experiments or other experimental protocols that require changing entire groups of users' experiences together~\cite{ugander2013graph, saveski2017detecting}, e.g., to understand equilibrium effects rather than individual effects. Our approach also cannot easily access the user's full inventory unless the experimenter implements a separate application layer to acquire it directly from the platform. In this paper, we will focus on changes to the feed items that have already been selected to be shown to the user.

\xhdr{Concrete Use-case} The effectiveness of the described framework was demonstrated in a large-scale experiment on X involving 1,256 participants~\cite{piccardi2025reranking}. Over a 10-day period leading up to the 2024 US presidential election, participants' feeds were reranked to evaluate the impact of content exposure on affective polarization and emotional responses. Participants' posts were processed by a backend system, initially filtered for political content using a custom BERT classifier. Political posts were then scored by GPT using well-established social science constructs to estimate each post’s anti-democratic content. Posts were subsequently up-ranked or down-ranked according to the assigned experimental condition. Compared to a control group whose feeds remained unaltered, participants exposed to fewer posts that contained anti-democratic content reported decreased affective polarization and a temporary reduction in anger and sadness. Conversely, increased exposure to anti-democratic content posts resulted in heightened affective polarization and negative emotions. Given the urgency of independent research on the impact of social media platforms~\cite{allen2025platform}, here we summarize lessons learned from implementing this experiment to support the community in designing similar studies.

\subsection{Key concepts} In this section, we describe the key concepts required to understand this guide.

\xhdr{Browser extensions} This guide focuses on modifying social media feeds using Browser extensions. Browser or web extensions are add-ons typically available via the browser-specific store, designed to enhance the browser's functionality. They support developers in customizing the browser's behavior and modifying websites' content. This capability is advantageous for developing custom social media feed interventions, enabling design-control trials or A/B tests. From the perspective of the experimental participants, the only technical prerequisite is to install the extension and accept the permissions.
With authorization, browser extensions can modify the pages of interest to achieve the desired intervention. These modifications to the web pages range from updating the content and its visual appearance to the website's behavior, such as the response to user interactions. Section \ref{sec:implementation} provides an overview of implementation details.

\xhdr{Platform Inventory} In this guide, we refer to the full database of candidate posts used by the feed algorithm as the participant's \textit{inventory}. This is a crucial concept because, depending on the platform of interest, the level of access to the candidate set can enable or limit some interventions.
When the feed is organized chronologically, the algorithm can sort the posts of relevant users (e.g., friends or followed users) by time and return the most recent ones. When the feed is algorithmically curated, the algorithm must evaluate a large set of posts from users within and outside the focal user's first-degree social network. This process typically relies on multiple steps. For instance, Twitter/X's ``For you'' algorithm, as described in the open-sourced implementation~\cite{twitter_alg}, first selects content using a high-recall retrieval system that scans a large amount of data, which is then scored and ranked by a more refined model.

\section{Feed Ranking}
\label{sec:interventions}

This section summarizes the potential of browser extensions to customize social media feeds. The reranking can be a complete shuffle of the feed based on a new objective or more focused, for example, up-ranking or down-ranking content that the researchers or users~\cite{kolluri2025alexandria} believe should receive more or less exposure.
In this guide, we focus on these two use cases, up-ranking and down-ranking, where the content of interest is made easier or harder to reach, respectively, but these recommendations can be adapted for different types of rank adaptations. These edits can be achieved by combining up-ranking and down-ranking operations, which constitute the building blocks of a feed intervention. 

A key functionality of the system handling the experiment is identifying relevant content and scoring it according to the metric of interest. Depending on the research goal, this capability can be achieved at different levels of complexity, ranging from simple keyword matching to more complex logic that relies on external AI services, such as large language models (LLMs). For example, feed posts can be scored and rearranged to make hostile political content harder to reach, or to make posts that reflect positive emotions easier to reach.

These manipulations offer a unique opportunity to investigate the causal effects of variations of the rank objective in a realistic scenario. We summarise the challenges and limitations of both implementations.
Additionally, a browser extension can edit post content. We conclude this section with a discussion on content editing and how a browser extension can modify text or social metrics.  \Sectionref{sec:implementation} offers a detailed description of how these manipulations can be accomplished in practice.

\subsection{Down-ranking}
The ability to down-rank content allows researchers to investigate alternative moderation strategies. Potentially harmful content can be dynamically penalized in the ranking, making it available only when the user is interested in consuming more content and scrolling further down their feed. Given the criteria for down-ranking content, such interventions are generally easy to implement. The extension should select the posts that must be penalized and estimate how far down the feed each item should be down-ranked. The new position can be determined based on a fixed offset or based on the score of the content. For example, researchers may decide to push all down-ranked posts 100 positions lower in the feed or use a content-based offset to ensure that very harmful posts receive less exposure. In practice, the extension should keep track of the number of posts the user has consumed and insert a post when the browser's viewport reaches the desired position. This step can be easily achieved on most platforms by inserting the same content that was removed in the previous position.
This intervention offers valuable insights into the impact of ranking. It can provide concrete recommendations for the social media platform on how to adapt its algorithm to penalize problematic content and achieve the desired outcome. This intervention allows testing interventions that can reduce exposure to some content without entirely removing the recommended posts---a topic that may be sensitive for those opposing moderation strategies.

As an alternative, an extreme form of down-ranking is the complete removal of specific content. This logic, similar to how ad blockers work, operates on the definition of an exclusion logic or metric that guides the extension in identifying the posts that should be removed from the user's feed. This intervention can provide valuable insights into the impact of filtering out posts with particular characteristics. This approach allows researchers to investigate questions such as the impact of stricter moderation methods that reduce content with the potential to polarize communities or emotionally affect people, and the impact of alternative feed designs, such as removing all posts with videos or images.

Finally, when measuring the effect of down-ranking or removing content, it may be important to consider what the user was shown in place of the removed content. To answer these questions, we recommend keeping track of the posts the participant was exposed to during the study. 

\subsection{Up-ranking}
Without access to the full platform inventory, up-ranking content in the feed poses more challenges than down-ranking it. The content returned by the server is obtained by scoring the inventory, which includes the entire candidate set of potential posts available for display. Without the full set or a meaningful approximation, the intervention can operate only on the pre-selected content delivered to the user, representing the top items the curation algorithm has already prefiltered. 
For example, when the platform's algorithm selects content based on engagement, the posts available for reranking may consist only of high-engagement posts. Access to the full inventory could significantly expand the dynamic range of content available for reranking, potentially showing a more prominent effect of an intervention. However, compiling the full or even partial inventory may require more active support from the platforms or data access agreements, which may be difficult or expensive to set up. Possible approaches include up-ranking posts loaded by the website but not visible in the browser's viewport, pre-fetching content with the extension by simulating scrolling, or inserting entirely new posts from the platform inventory obtained in other ways.

Inserting new posts provides a way to customize the feed and expose users to content not selected by the curation algorithm. This intervention can help researchers investigate research questions that rely on adding content blended into the original feed. Some examples include adding more content that promotes positive emotions, adding out-party posts to counter filter bubbles, or adding posts on a specific topic to shift the feed's topic distribution. This strategy is particularly helpful when the type of posts the researchers are interested in investigating is rare, and the goal is to ensure the feed has enough relevant content.

From a technical standpoint, this approach requires careful consideration. The inserted posts can be sourced in three ways: (i)~generating entirely new posts that fit the requirements of the experiment, (ii)~monitoring posts by a curated list of accounts that are likely to post suitable content, and (iii)~transferring a post that matches these requirements from another feed. The first approach may present challenges because it requires replicating a fully functional social media post, including user interactions, such as likes, comments, and shares. This method may allow full customization but demands careful attention to detail to ensure the post supports real-world user interaction. Failing to recreate all these interactions accurately (i.e., the share button not working) may disrupt the user experience and compromise the study's validity by influencing user behavior in unanticipated ways.

The second approach is to monitor a set of public accounts likely to post content that meets the intervention's requirements. This method requires running a background process that continuously collects and scores the content posted by the selected accounts. There are two challenges with this approach: (i)~the content that individual users post varies, and curating a list of suitable accounts effectively may be challenging for some interventions; and (ii)~the data collection may require additional data access, which may be prohibitively expensive. 

The third approach is to transfer content from existing feeds. This approach offers a more pragmatic alternative, allowing researchers to utilize authentic posts. This 
technique has a downside: researchers must implement a mechanism to find or produce posts that qualify for the intended intervention, especially without a full inventory. This can be done by drawing the post from a dedicated account, using posts from the same user previously removed, or monitoring the feeds of other participants in the experiment. In the former case, this method requires researchers to ensure that the content selected for transfer respects user privacy and does not inadvertently expose private or sensitive information. Finally, researchers employing this intervention must reflect on the potential impact of context when transferring posts between feeds, as the relevance and reception of a post may vary significantly based on its original context.

\subsection{Content Editing} In addition to reranking the content of the feed, researchers can operate directly on individual posts. This intervention can involve direct manipulation of the post, such as editing the visible social metrics (e.g., likes, comments, and shares), the appearance of the post (e.g., making some aspects of the post prominent), changing the attachments (e.g., replacing a link or a picture), or modifying the text. These interventions may be important in addressing research questions on how the platform's design or content can affect the users. For example, the tone of existing posts can be subtly reframed to reflect a more positive sentiment or mitigate aggressive language. This intervention can allow researchers to explore how language choices in social media posts may influence user behavior, including interactions, affective experiences, discourse patterns, misinformation, or hate speech. Depending on the complexity of the desired edits, the modification can be achieved with simple dictionary substitution or complete reframing, e.g., leveraging LLMs to generate high-quality and contextual text.

\xhdr{Summary} This section summarises the opportunities and challenges associated with down-ranking, up-ranking, and editing feed content. Down-ranking is generally the simplest intervention, as it involves operating on content already available in the feed, whereas up-ranking, in the absence of access to the complete inventory, may require a more careful design to identify or generate posts to up-rank. Content editing can be combined with the previous methods and represents an effective strategy for testing the impact of social metrics, language, or post appearance.
Furthermore, thanks to the capabilities of browser extensions, these modifications can be combined with additional interventions. Some examples include extending the platform's functionality with widgets like the Twitter/X Community Notes labels, or adding visual indicators for certain post characteristics, such as a warning about language toxicity. These interventions can be combined to achieve more complex experimental designs that align with specific research goals.

Finally, all these interventions require researchers to pay attention to the distribution of posts in the feed of the target population, which may have repercussions for the recruitment strategy. For example, manipulating a single post may produce minimal exposure to the treatment condition, and the effect of the intervention may be too small to detect~\cite{guess2021experiments}. This problem is particularly impactful in situations when users interact with only a small fraction of the content~\cite{wang2016measuring}, typically at the top of the feed~\cite{bakshy2015exposure} or when the types of posts that the researchers intend to edit are rare.

\section{Measurement}

In this section, we describe a set of approaches for measuring the impact of the feed intervention. We summarize three alternatives that enable different insights into the effect of the feed manipulation: momentary measures, longitudinal surveys, and engagement signals.

\subsection{Ecological Momentary Assessments}
When researchers are interested in measuring the momentary impact of a feed intervention, a suitable design is to add surveys directly to the feed. These in-feed surveys, inspired by the approach developed in psychology called Ecological Momentary Assessment~\cite{shiffman2008ecological}, allow researchers to capture real-time reactions directly in the context where the intervention is applied~\cite{backstrom2016serving}. In the case of social media, this strategy can elicit direct feedback on the intervention at the right time, as users scroll through their feeds. Researchers can ask highly contextualized questions such as ``Are you interested in this post?'' or ``How does this post make you feel?''. This method can be used to measure a momentary impact of the intervention that may be too short to be measured with long-term surveys, or that is hard to measure when the question is posed without context. The insertion of these surveys can be triggered by specific events, such as when a specific post is displayed, added at regular intervals, or shown at random times while the user consumes their feed.
The appearance of the survey widget may vary, from attention-grabbing designs such as popups or modal windows that can interrupt scrolling, to more subtle designs integrated as a special post in the regular feed. If integrated with the regular feed, one practical challenge of in-feed surveys is making them visible and clear enough for participants to know these questions are part of the study. If researchers require that the insertions get noticed, we recommend using colors and animations that stand out from the default design. The design must also be adapted to the different templates (e.g., dark mode) that the social media platform supports. Finally, when deciding where to place the in-feed survey, researchers must carefully consider the goals of their measurement. Participants answering the survey will be exposed to the content that precedes it in the feed, and might be influenced by other content visible in their viewport.

\subsection{Survey methodology}
Researchers interested in measuring the cumulative effects of the treatment can rely on a standard pre-post survey design. In such a setup, participants are presented with the same survey before (pre-) and after (post-) the experiment. Examples of attitudinal shifts that can be caused by feed interventions include changes in affective polarization, often measured using the feeling thermometer, or opinion change, typically measured with Likert-scale surveys. These surveys are assigned at the start and the end of a study and require a seamless integration into the experimental design. Pre-surveys intended to be completed at the beginning of the experiment can also be used to assign participants to specific experimental conditions,  to determine whether they qualify for enrollment, or to assess systematic differences between participants who continue with the study and those who drop out. Attention checks and validation questions can be valuable tools for identifying early participants who may be less attentive and contribute unreliable data.

A similar approach, inspired by the diary study technique~\cite{bolger2003diary}, consists of regular assessments throughout the experiment.  This method relies on administering surveys at regular intervals, e.g., daily or weekly. This design may be recommended for several reasons, such as measuring the progressive change in outcomes of interest or for a study design that requires participants to be exposed to different experimental conditions (e.g., a crossover or stepped-wedge design).

Implementing a strategy to administer surveys at regular intervals can present additional challenges, especially when the users are not online on the social media platform. Researchers may need to establish a notification system via a browser extension, implement regular messaging on the recruitment platform, or distribute a survey link through additional communication channels such as email or text messaging. This last approach requires participants' consent to collect potentially identifiable information, which we discuss in more detail in Section \ref{sec:notifications}. Another challenge of this setup is ensuring participants consume enough content between surveys to meaningfully measure the intervention's impact. One potential approach, valid for repeated assessments, is to invite participants to use social media and to deliver the survey only after they have spent enough time on the platform or have been exposed to a predefined number of posts.

\subsection{Engagement signals}
Implementing a feed experiment using a browser extension enables researchers to go beyond self-reported measures and obtain very granular behavioral data that can provide further insights into the intervention's effects. This includes metrics such as time spent, engagement (e.g., clicks, likes/favorites, reposts), navigation patterns, conversion rates, and other metrics~\cite{cunningham2024we} that are imperative to the business objectives of the platforms. Access to such measures also allows researchers to evaluate the trade-offs introduced by the proposed intervention. For instance, removing some content entirely (e.g., all political posts) may affect an outcome of interest (e.g., affective polarization) while also limiting disruption to engagement metrics, making adoption more attractive to social media companies.

Similar to the previous section, these measurement approaches can be combined to measure multiple effects and increase the understanding of the intervention at different time horizons.

\section{Implementing a Browser Extension}
\label{sec:implementation}

In this section, we focus on the implementation details of running social media feed experiments with a browser extension and summarize various technical challenges. We offer practical recommendations and release source code~\footnote{\url{https://github.com/StanfordHCI/FeedMonitor}} that can be used as a blueprint to implement feed experiments on social media. This section will focus on X as a use case, but the same technological approach can be applied to any platform. Our examples are based on Google Chrome but can easily be adapted to other browsers. The rest of this section assumes familiarity with key concepts of web architecture and extension development.

First, we provide implementation guidance on how to intercept and edit the responses from the server containing the feed. Then, we cover guidelines on participant recruitment and registration. Finally, we conclude with technical details on how to notify the user about specific events or deliver surveys.

\subsection{Feed interventions}
In this section, we present key details on implementing the browser extension's core functionality. Although these extensions can modify many aspects of the website experience, here we specifically focus on changes related to social media feeds. Browser extensions are typically written in languages supported by the browser, such as JavaScript, HTML, and CSS, and can use various methods to modify the website experience. A common strategy involves modifying the rendered page by manipulating its Document Object Model (DOM). This method has the advantage of simplifying the identification of the element involved in the desired modification, and it gives the browser extension access to the content after the frontend scripts have already processed it. 
However, despite this approach being the most straightforward, for some scenarios, such as augmenting the feed, intervening at the network level may be more effective. 

In this case, we recommend adapting  XMLHttpRequest to support the customization of the server requests. This technique offers the advantage of intercepting and modifying the incoming data before it is rendered on the page, minimizing the need for direct interaction with the page interface. Accessing the server's communication allows logging many client-side actions shared with the server (e.g., likes and shares) without adding a listener for each event. The downside is the need to gain a deeper understanding of the communication protocol between the browser and the server. Finally, alternative approaches, such as manipulating the in-memory data structure used to render the content, are possible, but depending on the front-end framework (e.g., React), they may introduce implementation-specific complications.

\xhdr{Implementation details} Pragmatically, to customize the behavior of the XMLHttpRequest native object, we need to run the overriding code as soon as the main page is loaded. Since behavior customization must happen within the scope of the social media page and not in the content script, we need to inject the modification script directly into the main page. This requirement necessitates a slight departure from the common pattern for developing browser extensions, as the script must be loaded in the \textit{Manifest} file as a resource rather than as a content script. Then the content script can inject the override code into the main page's scope.

The injected code overrides the \textit{open} and \textit{send} functions of XMLHttpRequest to adapt their behavior. The \textit{open} function can be modified to save properties of the connection, such as the current URL or the HTTP headers, while the \textit{send} function can be adapted to dynamically customize the response callback (\textit{onreadystatechange}) if the URL matches the endpoint of interest. In this function, researchers interested in fetching more data (i.e., to build a large pool of posts to rerank) can mimic server requests to simulate users scrolling through the feed.
The injected script can communicate with the browser extension's content script using the browser's message-passing features by broadcasting the server's raw response.
Depending on the intervention logic, when the feed needs to be modified, the customized XMLHttpRequest object can pause execution while waiting for a response from the content script. Once the modified feed is ready, the request can resume by replacing the server response with the updated content. The content can be manipulated on the client side or by calling an external backend. If the logic or model can be distributed with the extension, keeping the manipulation on the client may ensure more participant privacy. Moving the edit to the backend may be preferable in other cases, like when complex logic or large models, such as LLMs, are required. Still, in this case, researchers must pay special attention to data security practices and the potential implications of sharing the data with third-party services.

\xhdr{Latency considerations}
It is essential to ensure that scoring and reranking processes introduce minimal latency to avoid disruption of the user experience. 
Social media feeds are typically loaded from the server in discrete batches: for example, the X "For You" feed loads approximately 35 items per request. Latency therefore primarily affects the moment when a user lands on the website and the feed is empty. After the initial batch of posts is rendered, platforms generally use infinite scrolling and preload subsequent batches in the background as users consume the feed. As a result, any additional latency following the initial load is largely absorbed by the platform's existing loading mechanisms and is rarely perceptible during normal browsing.

This design substantially mitigates latency concerns because the intervention typically introduces a delay only once per session. In practice, latency requirements are therefore reduced to ensuring that the initial feed load remains within an acceptable one-time waiting threshold. Additionally, by operating within the platform's native loading pipeline, this approach enables inheriting the platform's strategies for managing network delays, including default loading indicators to preserve fidelity to the original user experience.

Previous research has identified several thresholds for tolerable waiting times. Studies indicate that delays of approximately 10 seconds or more often cause users to disengage or switch tasks~\cite{nielsen1993response}. Web-focused research suggests that waiting times up to 2 seconds are generally well tolerated~\cite{nah2004study}, with noticeable declines in user attitudes occurring beyond 4 seconds~\cite{galletta2004web}. Other studies have reported upper tolerable limits of up to 12 seconds~\cite{hoxmeier2000system}, depending on context and task characteristics.

Consistent with these findings, in our large-scale field study on X involving 1,256 participants~\cite{piccardi2025reranking}, we introduced an average one-time feed-loading delay of approximately 3 seconds. In a post-study survey assessing user experience, only 24 participants (1.9\%) mentioned noticing the additional latency, suggesting that delays of this magnitude are unlikely to significantly disrupt the browsing experience when they occur only during initial feed load.

When an intervention relies on server-side logic, latency arises from both network transfer and content processing. Network delay can be reduced by applying best practices in global server load balancing (GSLB). However, this optimization is often unnecessary if the majority of study participants are located within a single geographic region. In cases where the intervention can operate on the client side, network request time may be minimized by executing scripts or models directly in the browser, utilizing technologies such as WebGL or WebAssembly-based implementations. Conversely, backend performance becomes critical when the intervention is dependent on server-side logic. To maintain acceptable waiting times, implementations may require careful parallelization, particularly when using external services such as large language model APIs or using dedicated hardware, such as GPUs, for specialized models like fine-tuned BERT.

Finally, when designing an experiment, it is important to ensure that any latency is applied uniformly across experimental conditions. In particular, even in control conditions where the feed remains unchanged, the waiting time should be comparable. This can be achieved by executing the same intervention pipeline and discarding the modified output before returning it.

\subsection{Participant recruitment}
\label{sec:recruitment}

The recruitment strategy depends on many aspects, such as availability constraints, specific demographic requirements (e.g., partisans or users of specific age groups), or the length of the study.

Some experiments may aim to recruit a participant sample representative of the platform's user base, which requires adjusting the recruitment based on the platform's user composition~\cite{guess2021experiments}. For example, in 2023, Twitter was used by 22\% of the adult population, with a bias toward young, highly educated, and liberal users ~\cite{socialmediafacts}, while Facebook is more balanced on these properties but has a larger active user base of women.

Depending on the requirements and the recruitment platform, it may be desirable to first screen participants through a short survey. This preliminary screening may help exclude individuals who do not meet eligibility criteria that the recruitment platform cannot filter for, such as how frequently they use social media. This step is crucial to selecting the population of interest and ensuring that the study includes the desired participants to receive the treatment. Researchers investigating interventions on extreme or niche content need to ensure that the recruited participants are typically exposed to the posts of interest. For example, a study found that during the 2016 election, 80\% of the fake news on Twitter was consumed by only 1\% of the users~\cite{grinberg2019fake}.
If this step is necessary, we recommend keeping the survey short to reduce costs and compensate participants for their time, even if they do not qualify, in accordance with common practices.
Alternative methods include using ads directly on the platform of interest~\cite{zhang2020quota,rosenzweig2020survey}, direct messages~\cite{guess2021experiments}, or snowball sampling with other rewards, such as gift cards or prizes.

\subsection{Registration flow}

Once participants are recruited, the first challenge is to ensure a low-effort, scalable onboarding process. A confusing onboarding procedure can lead to high participant dropout and increase the risk of researchers being overloaded with requests for assistance.

Along with clear instructions, we recommend simplifying the process by designing a registration flow that eliminates common mistakes, such as the need for participants to copy and paste information manually. We also recommend that participants complete all the onboarding steps only by following clickable links to ensure a smooth user experience.

Since browser extension stores like the Chrome Web Store do not support passing parameters during installation (e.g., participant ID), this section presents a solution to address this issue. This approach requires a coordination service, which can be integrated with the extension backend. It allows tracking the user's registration to assign user-specific configuration to the extension, sourced from the recruitment platform or an initial survey.

\label{sec:registration}
\begin{figure*}[t]
    \includegraphics[width=0.95\linewidth]{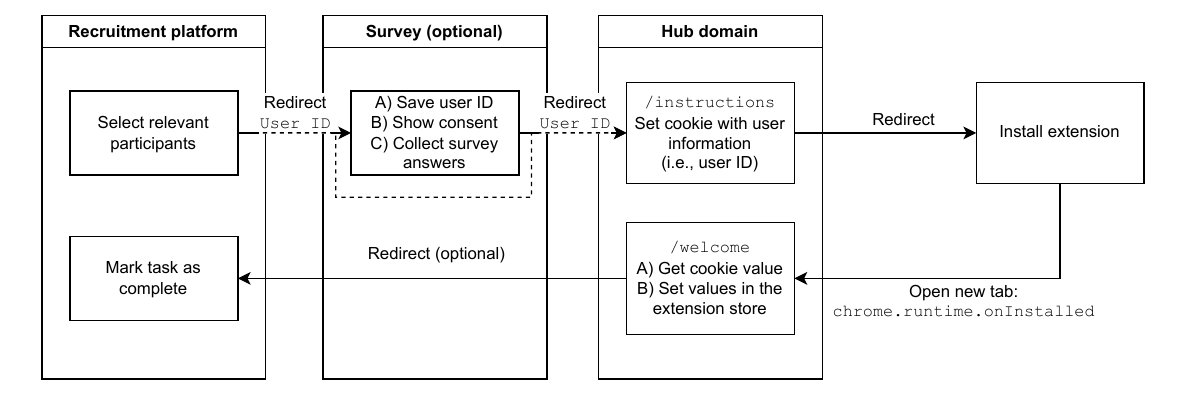}
    \caption{Diagram of the registration flow that allows for tracking the participant and for passing parameters to the extension. }
    \label{fig:onboarding_flow}
\end{figure*}

\Figref{fig:onboarding_flow} summarises the registration flow using this pattern. With this flow, the recruitment platform sends the user directly to the coordination service or via an intermediary survey, which could be a pre-survey used to collect additional information necessary to run the extension (e.g., political ideology, contact information, timezone). Major survey platforms like Qualtrics support transferring information after the survey completion by combining embedded fields and redirecting URLs. The role of the coordination service is to set a persistent entry, such as an HTTP cookie, containing the information the extension must access after installation. This page precedes the installation of the browser extension and may be used to summarize the instructions before redirecting the participant to the extension store page. Finally, after the extension is installed, opening a new tab to a page on the coordination server provides access to the previously saved HTTP cookie data, which can be used to set up the extension for the current participant.
This flow is simple and effective, but it is not the only approach that researchers can take. Alternative designs include integrating a custom login or implementing an OAuth flow.

\xhdr{Participant consent}
Running an experiment that augments users' feeds requires participant consent. Depending on the registration flow---with or without the initial survey---the participants should be presented with a description approved by an Institutional Review Board (IRB) or an equivalent record in other cases. It is important to show the consent form as early as possible and interrupt the installation if the participants do not accept the terms. Some participants may perceive installing a browser extension as too invasive. In addition to the consent, we recommend including in the instruction page a clear summary of what data is and what data is \textit{not} collected (e.g., ``we do not have access to your direct messages'').

\xhdr{Error recovery} Anything (and everything) could go wrong. Especially when the goal is to run a large-scale study, it is important to implement recovery points to minimize the researchers' workload of fixing errors. For example, a participant may start the survey using a browser or a computer different from the one they plan to use during the study. It is essential to implement recovery checkpoints so participants can recover the process. One option is to collect contact information, such as email addresses, and send a message when they reach the instructions page. The message can include a copy of the instructions and a link to the instructions page, with all required parameters in the URL. Alternatively, if researchers do not have access to participants' contact information---due to IRB restrictions or because no pre-survey is planned---this step can be automated via a message on the recruitment platform.

\subsection{Notifications}
\label{sec:notifications}
Throughout the study, researchers must establish communication channels with participants. Participants may need guidance, reminders, or feedback to ensure they engage in the study successfully. For example, in a longitudinal study where it is crucial to visit the social media platform regularly, the study design can use emails to remind participants that they are enrolled in the study or to acknowledge that their contributions have been recorded. Similarly, upon completion of the study, users may receive a post-survey, instructions to remove the extension, or debriefing notes. This step must be handled carefully to comply with Institutional Review Board (IRB) standards for data protection. 

Alternatively, user messages can be delivered directly on the platform by editing the social media page's DOM. The browser extension can add these messages as in-page banners or modal windows, limiting the need to collect user contact information. In the case of post-surveys showing a pop-up or in-page banner when participants visit the social media platform at the end of the study, researchers can prompt immediate and contextual responses, potentially improving the quality of the data collected.

\subsection{Privacy}
Browser extensions offer a great opportunity to test interventions that social media platforms would never try because of company priorities or because they do not align with the business objective. Nevertheless, even with greater freedom, researchers must still commit to ethical standards in designing experiments and handling data. Independent oversight, such as Institutional Review Boards (IRBs), should review and approve the study protocol to ensure it meets ethical standards. Participants must be aware of the potential risks and be informed about the data the browser extension collects. Designing a registration flow that requires participants to sign a consent form before collecting any data is essential. It is also crucial to inform participants that their participation is voluntary and that they can withdraw without penalty.
If the experiment collects users' data, it is essential to implement measures to ensure confidentiality and store the data in a secure database. Depending on where the study is conducted, researchers must comply with local regulations, such as the General Data Protection Regulation (GDPR) in the European Union.

\subsection{End of the experiment}
Unless specified in the consent form, at the end of the experiment, it is critical to implement an offboarding mechanism to ensure the browser extension stops collecting data and applying interventions.
We recommend that participants be guided to uninstall the extension or that the extension include a mechanism to turn itself off. To inform participants that the extension may still be installed, we recommend adding a clear visual signal, such as an overlay banner, with explicit instructions for uninstallation. This step ensures that the researchers do not collect data without the participant's consent and that the unused extension does not create potential conflicts with other studies the participant may want to join.

\section{Discussion}

\xhdr{Ethical considerations}
Given the central role of social media platforms in contemporary society, it is essential to maintain an independent and critical understanding of their societal impact. Previous research has argued that such inquiry may be ethically necessary to protect democratic processes~\cite{fiske2022twitter}.

Simultaneously, the approach outlined in this paper enables interventions that can significantly alter participants' information exposure. By inserting new posts into the feed or modifying existing content, these systems may influence users' emotions, attitudes, and perceptions. Consequently, researchers must rigorously evaluate the ethical implications of their interventions.

In addition to the privacy concerns previously addressed, researchers should ensure that the content presented to participants reflects a meaningful risk-benefit trade-off. Interventions must be designed to minimize potential harm while providing clear societal or scientific value. As these studies involve human participants and are conducted within real-world information environments, all interventions require review and approval by the appropriate Institutional Review Board (IRB) and must be implemented only with informed participant consent.

Finally, after the experiment ends, we encourage minimizing harm by following established ethical standards for participant debriefing, clearly explaining the study's purpose, and disclosing any manipulation or deception used in the design.

\xhdr{Limitations}
The proposed framework enables researchers to conduct field experiments directly on live social media platforms, thereby enhancing ecological validity. However, this approach has some limitations.

First, the implementation depends on the stability of platform APIs, which may change over time without warnings and disrupt the intervention. Although in our experience, the frontend APIs remain relatively stable for long periods, it is advisable for researchers to implement diagnostic and monitoring tools to ensure that the extension continues to function as intended, especially during extended deployments.

Second, researchers should consider potential feedback loops between the intervention and the platform's recommendation algorithm. Down-ranking specific types of content and reducing related engagement signals may prompt the algorithm to further limit recommendations of that content. In some instances, this effect may support the intervention's objectives by reinforcing the down-ranking of undesirable content. However, such algorithmic adaptation can complicate causal interpretation, so researchers should carefully assess how these dynamics may influence their conclusions.

Finally, the current browser extension implementation has been evaluated exclusively on desktop devices. This limitation may introduce biases in both the study population and observed behaviors. Participants may still access the platform on mobile devices, where the intervention is not applied, potentially diluting treatment exposure. This issue can be partially addressed by reminding participants to avoid using other devices during the study period. Additionally, researchers may ask participants whether they accessed the platform through unsupported devices, or estimate this by comparing the recorded user traces to their overall public activity (e.g., comparing posts liked through the extension with all posts the participant liked), and incorporate such information into their analysis. It is important to clearly communicate to participants that accessing the platform through other means will not affect their participation or compensation.

\xhdr{Beyond desktop}
Implementing our framework on mobile devices presents unique challenges. Unlike desktop browsers, modifying official mobile applications may violate the platform's terms of service. Additionally, security features such as certificate pinning prevent interception or modification of network traffic, even on rooted or jailbroken devices.

Stricter app store policies also prevent the distribution of unauthorized versions of major social media applications. Consequently, developing a generic and reproducible mobile solution remains difficult.
Researchers interested in extending this framework to mobile environments could consider developing custom applications that use embedded web views to load the mobile version of the platform. Because such custom apps offer greater control over the web view, they can inject largely the same JavaScript logic used in the desktop browser extension. Compared to browser-based extensions, an additional challenge lies in distribution. On Android, custom applications can typically be distributed as APK files outside the official app store. On iOS, however, distribution policies are significantly more restrictive. A possible approach is to use a beta distribution through TestFlight, though Apple's review process may still reject such applications.
Some mobile browsers, such as Microsoft Edge for mobile\footnote{\url{https://play.google.com/store/apps/details?id=com.microsoft.emmx.canary}}, offer limited browser extension support. However, at the time of writing, this feature is only available in beta versions.

\xhdr{Conclusion}
Social media platforms have recently become less collaborative in supporting independent academic research~\cite{allen2025platform}. This guide offers practical recommendations for designing field experiments that use browser extensions and do not require direct involvement from these platforms. 
Our recommendations are focused on social media feeds, but many aspects generalize beyond this setup. With this guide, we hope to foster a community of independent researchers who aim to investigate social media's impact on society and to contribute to the design of healthier online spaces.

\section*{Acknowledgements}
This work was supported in part by the National Science Foundation under awards IIS-2403433, IIS-2403434, IIS-2403435, and BCS-2214203; the Swiss National Science Foundation under award P500PT-206953; and a Hoffman-Yee grant from the Stanford Institute for Human-Centered Artificial Intelligence.

\bibliographystyle{ACM-Reference-Format}
\bibliography{references}

\end{document}

%% file: macros.tex
\usepackage{listings}
\usepackage{color}
\usepackage{xcolor}
\usepackage{framed}
\usepackage{amsthm}
\colorlet{punct}{red!60!black}
\definecolor{background}{HTML}{EEEEEE}
\definecolor{delim}{RGB}{20,105,176}
\colorlet{numb}{magenta!60!black}
\definecolor{lightgray}{rgb}{.9,.9,.9}
\definecolor{darkgray}{rgb}{.4,.4,.4}
\definecolor{purple}{rgb}{0.65, 0.12, 0.82}

\lstdefinelanguage{JavaScript}{
  keywords={typeof, new, true, false, catch, function, return, null, catch, switch, var, if, in, while, do, else, case, break},
  keywordstyle=\color{blue}\bfseries,
  ndkeywords={class, export, boolean, throw, implements, import, this},
  ndkeywordstyle=\color{darkgray}\bfseries,
  identifierstyle=\color{black},
  sensitive=false,
  comment=[l]{//},
  morecomment=[s]{/*}{*/},
  commentstyle=\color{purple}\ttfamily,
  stringstyle=\color{red}\ttfamily,
  morestring=[b]',
  morestring=[b]"
}

\lstdefinelanguage{json}{
    basicstyle=\footnotesize\ttfamily,
    numbers=left,
    numberstyle=\footnotesize,
    stepnumber=1,
    numbersep=8pt,
    showstringspaces=false,
    breaklines=true,
    frame=lines,
    literate=
     *{0}{{{\color{numb}0}}}{1}
      {1}{{{\color{numb}1}}}{1}
      {2}{{{\color{numb}2}}}{1}
      {3}{{{\color{numb}3}}}{1}
      {4}{{{\color{numb}4}}}{1}
      {5}{{{\color{numb}5}}}{1}
      {6}{{{\color{numb}6}}}{1}
      {7}{{{\color{numb}7}}}{1}
      {8}{{{\color{numb}8}}}{1}
      {9}{{{\color{numb}9}}}{1}
      {:}{{{\color{punct}{:}}}}{1}
      {,}{{{\color{punct}{,}}}}{1}
      {\{}{{{\color{delim}{\{}}}}{1}
      {\}}{{{\color{delim}{\}}}}}{1}
      {[}{{{\color{delim}{[}}}}{1}
      {]}{{{\color{delim}{]}}}}{1},
}

\newcommand{\Sectionref}[1]{Section~\ref{#1}}

\newcommand{\Figref}[1]{Fig.~\ref{#1}}

\colorlet{shadecolor}{blue!5}
\newtheoremstyle{mystyle}% name
  {3pt}% Space above
  {3pt}% Space below
  {}% Body font
  {}% Indent amount
  {\bfseries}% Theorem head font
  {.}% Punctuation after theorem head
  {.5em}% Space after theorem head
  {}% Theorem head spec (can be left empty, meaning ‘normal’)

\theoremstyle{mystyle}

\newcommand{\xhdr}[1]{\vspace{1.7mm}\noindent{{\bf #1.}}}